\documentclass[aps,pra,groupedaddress]{revtex4}
\usepackage{color,graphicx,bm,amssymb}
\begin{document}

%\preprint{}

\title{Coulomb corrections to bremsstrahlung in electric
field of heavy atom at high energies}

\author{R.N.Lee}
\email{R.N.Lee@inp.nsk.su}
\author{A.I. Milstein}
\email{A.I.Milstein@inp.nsk.su}
\author{V.M. Strakhovenko}
\email{V.M.Strakhovenko@inp.nsk.su}
\affiliation{Budker Institute of Nuclear
Physics, 630090 Novosibirsk, Russia}
\author{O.Ya. Schwarz}
\affiliation{Novosibirsk State University, 630090 Novosibirsk, Russia}

\date{\today}

\begin{abstract}
The differential and partially integrated cross sections are considered for
bremsstrahlung from high-energy electrons in atomic field with the exact
account of this field. The consideration exploits the quasiclassical electron
Green's function and wave functions in an external electric field. It is shown
that the Coulomb corrections to the differential cross section are very
susceptible to screening. Nevertheless, the Coulomb corrections to the cross
section summed up over the final-electron states are independent of screening
in the leading approximation over a small parameter $1/mr_{scr}$ ($r_{scr}$ is
a screening radius, $m$ is the electron mass, $\hbar=c=1$). Bremsstrahlung from
an electron beam of the finite size on heavy nucleus is considered as well.
Again, the Coulomb corrections to the differential probability are very
susceptible to the beam shape, while those to the probability integrated over
momentum transfer are independent of it, apart from the trivial factor, which
is the electron-beam density at zero impact parameter. For the Coulomb
corrections to the bremsstrahlung spectrum, the next-to-leading terms with
respect to the parameters $m/\varepsilon$ ($\varepsilon$ is the electron
energy) and $1/mr_{scr}$  are obtained.
\end{abstract} \pacs{} \maketitle

\section{Introduction}

Bremsstrahlung in the electric field of atoms is the fundamental QED process.
Its investigation, started in 30s, is important for various applications. In
the Born approximation both differential cross section and spectrum of the
bremsstrahlung have been obtained for arbitrary electron energies and atomic
form factors \cite{BH1934} (see also Ref. \cite{BLP1982}). High-energy
asymptotics of the bremsstrahlung cross section in a Coulomb field has been
studied in detail in Ref. \cite{BetheM1954} exactly in the parameter $Z\alpha$
($Z$ is the atomic number, $\alpha=1/137$ is the fine-structure constant). In
these papers the differential cross sections and the bremsstrahlung spectrum
have been obtained. For a screened Coulomb field, the high-energy asymptotics
of the differential cross section has been derived in Ref. \cite{OlsenMW1957}.
Influence of screening  on the spectrum has been studied in Refs.
\cite{Olsen1955,OlsenM1959}. For the spectrum, it turned out that screening is
essential only in the Born approximation. In other words, the Coulomb
corrections (CC) to the spectrum are not significantly modified by screening.
By definition, CC are the difference between the result obtained exactly in the
external field and that obtained in the Born approximation. In the recent paper
\cite{Olsen2003} it has been claimed that CC to the differential cross section
of the bremsstrahlung are also independent of screening.

In the present paper we investigate the bremsstrahlung cross section in the
electric field of a heavy atom. We assume that $\varepsilon\,,\,\varepsilon'\gg
m$, where $\varepsilon$ and $\varepsilon'$ are the initial and final electron
energies, respectively. In Sec. \ref{sec:CrossSection} we consider in detail
the differential cross section in the leading approximation, i.e., neglecting
corrections in the parameters $m/\varepsilon$ and $1/mr_{scr}$. In contrast to
the statement of Ref. \cite{Olsen2003}, screening may strongly modify CC to the
differential cross section. We demonstrate explicitly that this fact does not
contradict the final-state integration theorem \cite{Olsen1955} from which it
follows that CC to the spectrum are independent of screening. We also study the
influence of the electron beam finite size on CC. Again, CC to the differential
cross section are very sensitive to the shape of the electron beam, while the
spectrum is independent of it, except for a trivial factor. In Sec.
\ref{sec:correction_to_spectrum} we consider corrections to CC in the spectrum.
It turns out that, in the first non-vanishing order, they enter the spectrum as
a sum of two terms. The first term is proportional to $m/\varepsilon$ and is
independent of screening. The second term is small in the parameter
$1/mr_{scr}$ and is independent of the energy.

Our approach is based on the use of the quasiclassical Green's function and the
quasiclassical wave function of the electron in an external field. Earlier this
method was successfully applied to the investigation of the photoproduction
process at high energy \cite{LMS2003a,LMS2004}.

\section{Differential cross section}\label{sec:CrossSection}

The cross section of the electron bremsstrahlung in the external field has the
form
\begin{equation}\label{eq:cs}
d\sigma^\gamma=\frac{\alpha}{(2\pi)^4\omega}\,d\bm{p'}\,d\bm{k}\,\delta
(\varepsilon-\varepsilon'-\omega)|M|^{2}\,,
\end{equation}
where $\bm{k}$ is the photon momentum, $\bm{p}$ and $\bm{p}'$ are the initial
and final electron momenta, respectively, $\omega=|\bm{k}|$,
$\varepsilon=\varepsilon_{p}=\sqrt{\bm{p}^2+m^2}$, and
$\varepsilon'=\varepsilon_{p'}$ . The matrix element $M$ has the form
\begin{equation}\label{eq:BS_amplitude_via_WF}
M=\int d\bm{r}\, \mbox{e}^{-i\bm{k}\cdot\bm{ r}}\bar\psi_{P'}^{(out)}(\bm{r})
\hat{e}^* \psi_{P}^{(in)}(\bm{r})\,.
\end{equation}
Here $\psi_{P}^{(in)}$ and $\psi_{P}^{(out)}$ are the wave functions of the
\emph{in-} and \emph{out-}state of the electron in an external field,
containing in their asymptotics the diverging and converging spherical waves,
respectively, and the plain wave with 4-momentum $P$;
$\hat{e}^*=e_{\mu}^*\gamma^\mu$, $e_{\mu}$ is the photon polarization 4-vector,
$\gamma^\mu$ are the Dirac matrices.

In \cite{LMS2000} the quasiclassical wave function of electron in arbitrary
localized potential has been found with the first correction in $m/\varepsilon$
taken into account. For the calculation of bremsstrahlung and $e^+e^-$
photoproduction cross section in the leading approximation one can use the
following form of the wave function \cite{LMS2000}
\begin{eqnarray}\label{eq:WF_with_correction}
\psi^{(in,\, out)}_P(\bm{r})&=& \pm\int \frac{d\bm{q}}{i\pi}
 \exp\left[i\bm{p}\cdot\bm{r}\pm iq^2\mp i\lambda \int_0^\infty dx V(\bm{r}_x)
\right]\left\{ 1 \mp \frac 1{2p}\int\limits_0^\infty dx\,
\bm{\alpha\cdot\nabla} V(\bm{r}_x) \right\} u_{P}\ ,
\nonumber\\
\bm{r}_x&=&\bm{r}\mp x \bm n +\bm{q} \sqrt{2|\bm r\cdot\bm n|/p}\,,\quad
\lambda=\mathrm{sgn}\ P^0\,,\quad \bm n=\bm{p}/p\,.
\end{eqnarray}
In this formula $\bm q$ is a two-dimensional vector lying in the plane
perpendicular to $\bm p$, the upper sign corresponds to $\psi^{(in)}_P$, $u_P$
is the conventional Dirac spinor. We remind one that the wave function
$\psi_{(-\varepsilon_p,-\bm p)}^{(in)}$ corresponds to the positron in the
final state with the 4-momentum $(\varepsilon_p,\bm p)$. For a Coulomb field,
the wave function (\ref{eq:WF_with_correction}) coincides with the usual
Furry-Sommerfeld-Maue wave function. When the angles between $\bm p$ and $\bm
r$ in $\psi^{(in)}_P(\bm{r})$,  and between $\bm p$ and $-\bm r$ in
$\psi^{(out)}_P(\bm{r})$ are not small, it is possible to replace $\bm r_x$ in
Eq. (\ref{eq:WF_with_correction}) by $\bm{R}_{x}=\bm{r}\mp x \bm{n}$. Then the
integral over $\bm q$ can be taken, and we come to the conventional eikonal
wave function
\begin{eqnarray}\label{eq:WF_eik_with_correction}
\psi^{(in,\,out)}_{P,\,eik}(\bm{r})&=&
 \exp\left[i\bm{p}\cdot\bm{r}\mp i\lambda \int_0^\infty dx V(\bm{R}_{x})
\right]\left\{ 1 \mp \frac 1{2p}\int\limits_0^\infty dx\,
\bm{\alpha\cdot\nabla} V(\bm{R}_{x}) \right\} u_{P}\,.
\end{eqnarray}

We direct the $z$-axis along the vector $\bm \nu =\bm{k}/\omega$, then $\bm
r=z\bm \nu +\bm \rho$. In this frame the polar angles of $\bm{p}$ and $\bm{p}'$
are small. We split the region of integration in Eq.
(\ref{eq:BS_amplitude_via_WF}) into two: $z>0$ and $z<0$. The corresponding
contributions to $M$ are denoted as $M_+$ and $M_-$ so that $M=M_++M_-$. For
$z>0$ the function $\psi_{p'}^{(out)}(\bm{r})$ has the eikonal form and we
obtain for $M_+$
\begin{eqnarray}\label{eq:M_+}
M_+&=&\int\limits_{z>0}d\bm{r}\int \frac{d\bm q}{i\pi}
  \exp\left\{i\bm q^2-i\bm{\Delta}\cdot\bm{r}
  -i\int\limits_0^\infty dx[V(\bm r-\bm n x+\bm q\sqrt{2z/p})+V(\bm r+\bm n' x)]
  \right\}
\nonumber\\
&&\times\bar u_{p'} \left[\hat e^*
  -\frac1{2p}\int\limits_0^\infty dx\hat e^*\bm{\alpha}
  \cdot\bm{\nabla}V(\bm r-\bm n x+\bm q\sqrt{2z/p})
  -\frac1{2p'}\int\limits_0^\infty dx\bm{\alpha}
  \cdot\bm{\nabla}V(\bm r+\bm n' x)\hat e^*\right]u_p\,,
\end{eqnarray}
where $\bm\Delta=\bm p'+\bm k-\bm p$ is the momentum transfer.

 In Eq. (\ref{eq:M_+}) we have
replaced $\sqrt{2|\bm r\cdot\bm n|/p}$ in the definition of $\bm r_x$ in Eq.
(\ref{eq:WF_with_correction}) by $\sqrt{2z/p}$. It is easy to see that within
our accuracy we can also replace in Eq. (\ref{eq:M_+}) the quantity $V(\bm
r+\bm n' x)$ by $V(\bm r+\bm n' x+\bm q\sqrt{2z/p})$ and consider the vector
$\bm q$ to be perpendicular to $z$-axis. After that we shift $\bm\rho\to
\bm\rho-\bm q\sqrt{2z/p}$ and take the integral over $\bm q$. We obtain
\begin{eqnarray}\label{eq:M_+_no_q}
M_+&=&\int\limits_{z>0}d\bm{r}\exp\left\{-i\frac{z}{2p}{\Delta}_\perp^2-i\bm{\Delta}\cdot\bm{r}
  -i\int\limits_0^\infty dx[V(\bm r-\bm n x)+V(\bm r+\bm n' x)]\right\}
\nonumber\\
&&\times\bar u_{p'}\left[\hat e^*
  -\frac1{2p}\int\limits_0^\infty dx\hat e^*\bm{\alpha}
  \cdot\bm{\nabla}V(\bm r-\bm n x)
  -\frac1{2p'}\int\limits_0^\infty dx\bm{\alpha}
  \cdot\bm{\nabla}V(\bm r+\bm n' x)\hat e^*\right]u_p
\end{eqnarray}

In the same way, we obtain for $M_-$:

\begin{eqnarray}\label{eq:M_-_no_q}
M_-&=&\int\limits_{z<0}d\bm{r}\exp\left\{i\frac{z}{2p'}{\Delta}_\perp^2-i\bm{\Delta}\cdot\bm{r}
  -i\int\limits_0^\infty dx[V(\bm r-\bm n x)+V(\bm r+\bm n' x)]\right\}
\nonumber\\
&&\times\bar u_{p'}\left[\hat e^*
  -\frac1{2p}\int\limits_0^\infty dx\hat e^*\bm{\alpha}
  \cdot\bm{\nabla}V(\bm r-\bm n x)
  -\frac1{2p'}\int\limits_0^\infty dx\bm{\alpha}
  \cdot\bm{\nabla}V(\bm r+\bm n' x)\hat e^*\right]u_p
\end{eqnarray}

There are two overlapping regions of the momentum transfer $\Delta$,
\begin{equation}
\mathrm{I.~}\Delta\ll \frac{m\omega}\varepsilon\,,\quad \mathrm{II.~}\Delta\gg
\Delta_{min}=\frac{m^2\omega}{2\varepsilon\varepsilon'}\,.
\end{equation}

In the first region we can neglect the terms $\propto {\Delta}_\perp^2$ in the
exponents in Eqs. (\ref{eq:M_+_no_q}) and (\ref{eq:M_-_no_q}). Then the sum
$M=M_++M_-$ reads
\begin{eqnarray}\label{eq:M}
M&=&\int d\bm{r}\exp\left\{-i\bm{\Delta}\cdot\bm{r}
  -i\int\limits_0^\infty dx[V(\bm r-\bm n x)+V(\bm r+\bm n' x)]\right\}
\nonumber\\
&&\times\bar u_{p'}\left[\hat e^*
  -\frac1{2p}\int\limits_0^\infty dx\hat e^*\bm{\alpha}
  \cdot\bm{\nabla}V(\bm r-\bm n x)
  -\frac1{2p'}\int\limits_0^\infty dx\bm{\alpha}
  \cdot\bm{\nabla}V(\bm r+\bm n' x)\hat e^*\right]u_p
\end{eqnarray}

We can make the replacement $\bm n,\bm n'\to\bm \nu $ in the preexponent in Eq.
(\ref{eq:M}). In the exponent we should take into account the linear term of
expansion of the integral with respect to $\bm n-\bm \nu $ and $\bm n'-\bm \nu
$. As a result we have

\begin{eqnarray}\label{eq:M1}
&&M=\int d\bm{r}\exp\left[-i\bm{\Delta}\cdot\bm{r}
  -i\chi(\bm\rho) \right]
\nonumber\\
&&\times  \int\limits_0^\infty dy
   \bar u_{p'}\biggl[
   \hat e^* [i\, y(\bm n-\bm \nu )-\bm{\alpha}/2p]
   \cdot\bm\nabla V(\bm r-\bm \nu y)
   +[-i\, y(\bm n'-\bm \nu ) -\bm{\alpha}/2p']\cdot
   \bm\nabla V(\bm r+\bm \nu y)   \hat e^*
\biggr] u_p\,,\nonumber\\
&& \chi(\bm{\rho})=\int_{-\infty}^{\infty} dz V(\bm{r}) \,.
\end{eqnarray}

In the arguments of $V(\bm r\pm\bm \nu y)$ we make the substitutions $z\to z\mp
y$. After that we take the integral over $y$ and obtain
\begin{eqnarray}\label{eq:M_Small_Delta}
&&M=\bm{A}(\bm{\Delta})
 \cdot \left(
   \bar u_{p'}\biggl[
   \frac{(\bm n-\bm n')\hat e^* }{\Delta_z^2}
   -\frac{\hat e^*\bm{\alpha}}{2p\Delta_z}
   +\frac{\bm{\alpha}\hat e^*}{2p'\Delta_z}
\biggr] u_p\right)\,, \quad  \bm{A}(\bm{\Delta})=-i\int d\bm{r}
\exp[-i\bm{\Delta}\cdot \bm{r}-i
  \chi(\bm{\rho})]
  \bm{\nabla}_{\rho}V(\bm{r})\,.
\end{eqnarray}

Let us pass to the calculation of $M$ in the second region where $\Delta\gg
\Delta_{min}$. In Eq. (\ref{eq:M_+_no_q}) for $M_+$ we can replace $\bm n'\to
\bm n$ and $z\Delta_{\perp}^2/2p\to \tilde{z}\Delta_{\perp}^2/2p$, where
$\tilde{z}=\bm r\cdot\bm n$. Due to the smallness of the polar angle of $\bm n$
we can integrate in Eq. (\ref{eq:M_+_no_q}) over the half-space $\tilde{z}>0$.
After the integration over $\tilde{z}$ we obtain
\begin{eqnarray}\label{eq:M_+_Large_Delta}
M_+&=&-i\int d\bm{\rho}\exp\left[-i\bm{\Delta}\cdot\bm{\rho}
-i\chi(\bm\rho)\right]\, \frac{\bar u_{p'}\hat e^*
  \left[2p+\bm{\alpha}\cdot\bm{\Delta}_{\perp}\,\right]u_p}
  {2p\bm\Delta\cdot \bm n+\Delta_{\perp}^2}  \,.
\end{eqnarray}

The calculation of $M_-$ is performed quite similar. As a result we have
\begin{eqnarray}\label{eq:M_Large_Delta}
M&=&-i\int d\bm{\rho}\exp\left[-i\bm{\Delta}\cdot\bm{\rho}
-i\chi(\bm\rho)\right]\, \bar u_{p'}\left[\frac{\hat e^*
  \left(2p+\bm{\alpha}\cdot\bm{\Delta}_{\perp}\,\right)}
  {2p\bm\Delta\cdot \bm n+\Delta_{\perp}^2}-\frac{
  \left(2p'+\bm{\alpha}\cdot\bm{\Delta}_{\perp}\,\right)\hat e^*}
  {2p'\bm\Delta\cdot \bm n'-\Delta_{\perp}^2} \right]u_p  \,.
\end{eqnarray}

Now we can write the representation for $M$ which is valid in both regions
\begin{eqnarray}\label{eq:M_universal}
&&M= \frac{\varepsilon\varepsilon'}{\omega}\bm{A}(\bm{\Delta})
 \cdot\left\{
  \bar u_{p'}\biggl[
   -2\hat e^*\frac{\bm p_\perp+\bm p'_\perp}{\delta\delta'}
   +\frac{\hat e^*\bm\alpha}{\varepsilon \delta'}
   -\frac{\bm\alpha\hat e^*}{\varepsilon' \delta}
\biggr] u_p\right\}\,, \quad  \delta=m^2+\bm{p}_\perp^2\,, \quad
\delta'=m^2+{\bm{p}'}_\perp^2  \,.
\end{eqnarray}

This expression coincides within our accuracy with Eq. (\ref{eq:M_Small_Delta})
 in region I and with Eq. (\ref{eq:M_Large_Delta}) in region II. Using the
explicit form of the Dirac spinors, we finally obtain
\begin{eqnarray}\label{eq:M_final}
M&=&\frac{1}{2\delta\delta'} \bm{A}(\bm{\Delta}) \cdot
\left\{\varphi'^{\dag}\biggl[
  (\bm{p}_\perp+\bm{p}_\perp')
  \Bigl(\frac{\varepsilon+\varepsilon'}{\omega}\bm{e}^*\cdot(\bm{p}_\perp+\bm{p}_\perp')
  -i[\bm{\sigma}\times\bm e^*]\cdot(\bm{p}_\perp+\bm{p}_\perp')\right.\nonumber\\
&&
  \left.+2im [\bm\sigma\times\bm e^*]_z\Bigr)
  -(\delta+\delta')\Bigl(\frac{\varepsilon+\varepsilon'}{\omega}\bm{e}^*
  -i[\bm{\sigma}\times\bm e^*]_{\perp}\Bigl)
  \biggr]\varphi\right\}\,.
\end{eqnarray}

This expression is in agreement with that obtained in \cite{OlsenMW1957} by
another method. We emphasize that the potential enters the amplitude
(\ref{eq:M_final}) only via $\bm A(\bm\Delta)$.

\subsection{CC to the differential cross section in a screened Coulomb potential}\label{sec:CCscr}

Let us discuss CC to the differential cross section of bremsstrahlung. We
remind that these corrections are the difference between the exact (in the
external field strength) cross section and that obtained in the Born
approximation which is proportional to $[|\bm A(\bm \Delta)|^2-|\bm{A}_B(\bm
\Delta)|^2]$ with $\bm A(\bm \Delta)$ from Eq. (\ref{eq:M_Small_Delta}) and
\begin{equation}\label{eq:A_B}
 \bm{A}_B(\bm \Delta)=-i\int d\bm{r}
  \exp[-i\bm{\Delta}\cdot \bm{r}]
  \bm{\nabla}_{\rho}V(\bm{r})
  =\bm\Delta_\perp \int d\bm r
  \exp[-i\bm{\Delta}\cdot \bm{r}]    V(\bm{r}) \,.
\end{equation}

\begin{figure}
 \centering\setlength{\unitlength}{0.1cm}
\begin{picture}(85,80)
 \put(40,0){\makebox(0,0)[t]{$\zeta$}}
 \put(-6,30){\rotatebox[origin=c]{90}{$\Delta_\perp dR/d\Delta_\perp$}}
\put(0,0){\includegraphics[width=80\unitlength,keepaspectratio=true]{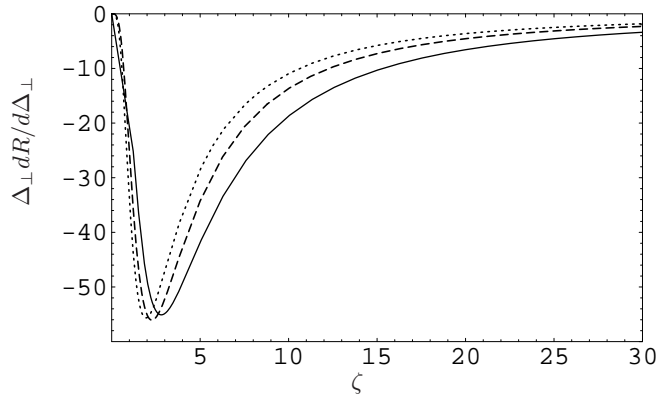}}
\end{picture}
\caption{The quantity $\Delta_\perp\,dR/d\Delta_\perp$ as a function of $\zeta$
for $Z=80$ and $\gamma=1$ (solid curve), $\gamma=0.5$ (dashed curve), and
$\gamma=0.01$ (dotted curve). The variable $\zeta$ is defined in Eq.
(\ref{eq:ydR_dy})}\label{fig:Delta_dependence}
\end{figure}

The screening modifies the Coulomb potential of the nucleus at distances
$r_{scr}\gg \lambda_C=1/m$. In the region $\Delta\gg
\mathrm{max}(\Delta_{min},\,r_{scr}^{-1})$ the quantities $\bm A(\bm\Delta)$
and $\bm A_B(\bm\Delta)$ are of the form
\begin{equation}
\bm{A}(\bm\Delta)=\bm{A}_B(\bm\Delta)\frac{\Gamma(1-iZ\alpha)}{\Gamma(1+iZ\alpha)}
\left(\frac{4}{\Delta_\perp^2}\right)^{-iZ\alpha}=-\bm\Delta_{\perp}\,\pi
Z\alpha \frac{\Gamma(1-iZ\alpha)}{\Gamma(1+iZ\alpha)}
\left(\frac{4}{\Delta_\perp^2}\right)^{1-iZ\alpha}\,.
\end{equation}
Thus $|A(\bm\Delta)|^2=|A_B(\bm\Delta)|^2$ for $\Delta\gg
\mathrm{max}(\Delta_{min},\,r_{scr}^{-1})$ and CC  to the differential cross
section vanish in this region in the leading approximation. Therefore, CC are
important only in the region $\Delta\lesssim
\mathrm{max}(\Delta_{min},\,r_{scr}^{-1})\ll m$. In this region we can use Eq.
(\ref{eq:M_Small_Delta}) for the matrix element. Substituting Eq.
(\ref{eq:M_Small_Delta}) in Eq. (\ref{eq:cs}) and using the relation
$d\Omega_{\bm p'}d\Omega_{\bm k}=d\phi d\bm\Delta_\perp
d\Delta_z/(\omega\varepsilon\varepsilon')$ we obtain for CC after integration
over the azimuthal angle $\phi$ and summation over polarizations
\begin{eqnarray}\label{eq:d4Sigma}
d\sigma_C^\gamma &=&\frac{\alpha d\omega d\bm\Delta_\perp
d\Delta_z}{16\pi^3\varepsilon^3\varepsilon'\Delta_z^2}
 \left[
 \varepsilon^2+\varepsilon'^2
 +2\frac{m^2\omega}{\Delta_z}
 +\frac{m^4\omega^2}{\varepsilon\varepsilon'\Delta_z^2}
 \right] \left[|\bm
A(\bm \Delta)|^2-|\bm{A}_B(\bm \Delta)|^2\right]   \,.
\end{eqnarray}
Note that in this formula we can assume that the $z$-axis is directed along the
vector $\bm p$. Then $\Delta_z$ is negative and $|\Delta_z|\geqslant
\Delta_{min}=m^2\omega/2\varepsilon\varepsilon'$.  The potential $V(\bm r)$ and
the transverse momentum transfer $\bm\Delta_\perp$ enter Eq. (\ref{eq:d4Sigma})
only as the factor $dR$,
\begin{equation}\label{eq:dR}
d R= d\bm\Delta_\perp\left[|\bm{A}(\bm \Delta)|^2-|\bm{A}_B(\bm
\Delta)|^2\right]\,.
\end{equation}
It follows from the definition of $\bm{A}(\bm \Delta)$ that for $r_{scr}\gg
|\Delta_z|^{-1}$ screening can be neglected. However, it is obvious from Eq.
(\ref{eq:dR}) that screening drastically modifies the
$\bm\Delta_\perp$-dependence of the differential cross section for
$r_{scr}\lesssim |\Delta_z|^{-1}$. We illustrate this statement on the example
of the Yukawa potential $V(r)=-Z\alpha\exp[-\beta r]/r$. After the
straightforward calculation we have
\begin{eqnarray}\label{eq:ydR_dy}
&&\Delta_\perp\,\frac{d R}{d\Delta_\perp}= 32\pi^3(Z\alpha)^2 \left[
 \zeta^2\left|\int_0^\infty dx\,x J_1(x \zeta) K_1(x)\exp[2iZ\alpha K_0(\gamma x)]\right|^2
 -\frac{\zeta^4}{(1+\zeta^2)^2}
 \right]\,,\nonumber\\
&&\zeta=\frac{\Delta_\perp}{\sqrt{\Delta_z^2+\beta^2}}\,,\quad
 \gamma=\frac{\beta}{\sqrt{\Delta_z^2+\beta^2}} \,.
\end{eqnarray}
We emphasize that $\Delta_\perp$ enters the right-hand side of Eq.
(\ref{eq:ydR_dy}) only via the variable $\zeta$, so that
$\sqrt{\Delta_z^2+\beta^2}$  is the characteristic scale of the distribution
(\ref{eq:ydR_dy}). For $\beta\gg |\Delta_z|$ this scale is entirely determined
by the radius of screening $r_{scr}=\beta^{-1}$. In this case the
$\Delta_\perp$-distribution is much wider than that in the absence of
screening. Therefore we conclude that, in contrast to the statement in Ref.
\cite{Olsen2003}, CC to the differential cross section strongly depend on
screening. Note that screening also affects the shape of the
$\Delta_\perp$-distribution (\ref{eq:ydR_dy}) via the parameter $\gamma$, which
varies from $0$ to $1$. In Fig. \ref{fig:Delta_dependence} we show the
dependence of $\Delta_\perp\,{d R}/{d\Delta_\perp}$ on scaling variable $\zeta$
for $Z=80$ and different values of the parameter $\gamma$.

Note that, in contrast to bremsstrahlung, CC  to the differential cross section
of $e^+e^-$ photoproduction in the atomic field are important only in the
region $\Delta_\perp\sim m$ where screening may be neglected
\cite{OlsenMW1957}.

\subsection{Integrated cross section}

It was shown in Ref. \cite{Olsen1955} that CC  to the cross section of
bremsstrahlung integrated over $\bm\Delta_\perp$ are independent of screening
in the leading approximation. The statement was based on the possibility to
obtain this cross section from that for $e^+e^-$ photoproduction. In this
subsection we perform the explicit integration of $d\sigma_C^\gamma$, Eq.
(\ref{eq:d4Sigma}), over $\bm\Delta_\perp$. We show that the strong influence
of screening on the shape of $d\sigma_C^\gamma$ does not contradict the
statement of Ref. \cite{Olsen1955}. Our consideration is quite similar to that
used in Ref. \cite{LM2000} at the calculation of CC  to the $e^+e^-$ pair
production in ultrarelativistic heavy-ion collisions.

 Let us consider the quantity $R$
\begin{equation}\label{eq:R}
R=\int d R=\int d\bm\Delta_\perp\left[|\bm{A}(\bm \Delta)|^2-|\bm{A}_B(\bm
\Delta)|^2\right]\,.
\end{equation}
This integral is converging due to the compensation in the square brackets in
Eq. (\ref{eq:R}). The main contribution to the integral comes from the region
$\Delta_\perp\lesssim \mathrm{max}(\Delta_{min},\,r_{scr}^{-1})$. Substituting
the integral representation for  $\bm{A}(\bm\Delta)$, Eq.
(\ref{eq:M_Small_Delta}), and for $\bm{A}_B(\bm\Delta)$, Eq. (\ref{eq:A_B}), to
Eq. (\ref{eq:R}), we have
\begin{equation}\label{eq:Rtriple}
R=\int d\bm\Delta_\perp\int\int d\bm r_1d\bm r_2
 \exp[i\bm\Delta\cdot(\bm r_1-\bm r_2)]
 \left\{  \exp[i\chi(\bm\rho_1)-i\chi(\bm\rho_2)]-1 \right\}
 [\bm{\nabla}_{1\perp}V(\bm{r}_1)]\cdot [\bm{\nabla}_{2\perp}V(\bm{r}_2)]\,.
\end{equation}

It is necessary to treat this repeated integral with care. If we naively change
the order of integration over $\bm\Delta_\perp$ and $\bm r_{1,2}$ and take the
integral over $\bm\Delta_\perp$ then we obtain $\delta(\bm\rho_1-\bm\rho_2)$.
After that the integration over $\bm \rho_1$ leads to zero result. This mistake
was made in Ref. \cite{OlsenMW1957} in the explicit check of the independence
of the integrated cross section on screening. Therefore, the proof of this
independence given in Ref. \cite{OlsenMW1957} and widely cited in textbooks is
not consistent. The correct integration of the cross section can be performed
as follows. Let us first integrate over the finite region, $\Delta_\perp<Q$, of
$\bm\Delta_\perp$. Then we can change the order of integration and first take
the integral over $\bm \Delta_\perp$. After that $R$ is obtained by taking the
limit $Q\to \infty$:
\begin{eqnarray}
R&=&\lim_{Q\to \infty}2\pi Q\int \int d\bm r_1d\bm r_2
 \frac{J_1(Q|\bm \rho_1-\bm \rho_2|)}{|\bm \rho_1-\bm \rho_2|}
 \exp[i\Delta_z(z_1-z_2)]\nonumber\\
 &&\times
 \left\{  \exp[i\chi(\bm\rho_1)-i\chi(\bm\rho_2)]-1 \right\}
 [\bm{\nabla}_{1\perp}V(\bm{r}_1)]\cdot [\bm{\nabla}_{2\perp}V(\bm{r}_2)]\,.
\end{eqnarray}
After the substitution $\bm r_{1,2}\to\bm r_{1,2}/Q$ we can pass to the limit
$Q\to\infty$ in the integrand using the asymptotics $V(\bm r)\to -Z\alpha/r$
and $\chi(\bm\rho)\to 2Z\alpha (\ln \rho+const)$ at $r\to 0$. Then we have
\begin{eqnarray}\label{eq:R1}
R&=&8\pi (Z\alpha)^2\int \int d\bm \rho_1d\bm \rho_2
 \frac{(\bm\rho_1\cdot\bm\rho_2)J_1(|\bm \rho_1-\bm \rho_2|)}
 {\rho_1^2\rho_2^2|\bm \rho_1-\bm \rho_2|}
 \left\{  \left(\frac{\rho_2}{\rho_1}\right)^{2iZ\alpha}-1 \right\}\nonumber\\
&=&
 -32 \pi^3(Z\alpha)^2[\mathrm{Re}\psi(1+iZ\alpha)+C]=
 -32 \pi^3(Z\alpha)^2f(Z\alpha)\,,
\end{eqnarray}
where $C$ is the Euler constant, and $\psi(x)= d\ln \Gamma(x)/dx$. Using this
formula and taking the integral over $\Delta_z$ from $-\infty$ to
$-\Delta_{min}$ in Eq. (\ref{eq:d4Sigma}), we reproduce the well-known result
obtained in Ref. \cite{BetheM1954}.

Thus, we come to a remarkable conclusion: CC to the integrated over $\bm
\Delta_\perp$ cross section are independent of screening, although the main
contribution to the integral comes from the region $\Delta_\perp\lesssim
\mathrm{max}(\Delta_{min},\,r_{scr}^{-1})$ where, for $\Delta_{min}\ll
r_{scr}^{-1}$, the differential cross section is essentially modified by
screening. We emphasize that this result is valid in the leading approximation
with respect to the parameters $m/\varepsilon\ll 1$ and $\lambda_C/r_{scr}\ll
1$. In the next section we show that in the limit $m/\varepsilon\to 0$ the
screening contributes to $d\sigma_C^\gamma/d\omega$ only as a correction in the
parameter $\lambda_C/r_{scr}$.

\subsection{Beam-size effect on CC }

It is interesting to consider the effect of the finite transverse size $b$ of
an electron beam on CC to bremsstrahlung in a Coulomb field of a heavy nucleus.
This consideration should be performed in terms of  the probability $dW$ rather
than the cross section. Similarly to the effect of screening, the finite beam
size can lead to the substantial modification  of CC to the differential
probability, $dW_C$ ,while CC to the probability integrated over $\bm\Delta$ is
a universal function. To illustrate this statement, let us consider
bremsstrahlung from the electron described in the initial state by the wave
function of the following form
\begin{equation}
\psi(\bm r)=\int d\Omega_{\bm p}\,h(\bm p) \psi_{P}^{(in)}(\bm r)\,.
\end{equation}
Here the function $h(\bm p)$ is peaked at $\bm p=\bm p_0$. If the width $\delta
p$ of the peak satisfies the condition $\delta p \ll \sqrt{\Delta_{min}
\varepsilon}\lesssim m$ then
\begin{equation}
\psi(\bm r)\approx\int d\Omega_{\bm p}\,h(\bm p)\exp[i(\bm p-\bm
p_0)\cdot\bm\rho] \psi_{P_0}^{(in)}(\bm r)=\phi(\bm\rho) \psi_{P_0}^{(in)}(\bm
r)\,,
\end{equation}
where the function $\phi(\bm\rho)$ is normalized as $\int d\bm\rho
|\phi(\bm\rho)|^2=1$ and has the width $b\gg 1/\sqrt{\Delta_{min}
\varepsilon}\gtrsim \lambda_C$. The quantity $dW_C$ is given by the right-hand
side of formula (\ref{eq:d4Sigma}) where the functions $\bm{A}(\bm{\Delta})$
and $\bm{A}_B(\bm{\Delta})$ are given by Eq. (\ref{eq:M_Small_Delta}) and Eq.
(\ref{eq:A_B}) with the additional factor $\phi(\bm\rho)$ in the integrands.
Substituting $V(r)=-Z\alpha/r$ we have

\begin{eqnarray}\label{eq:F_Coul}
\bm{A}(\bm{\Delta})&=&-2iZ\alpha \Delta_z \int d\bm{\rho}\phi(\bm\rho)
\exp[-i\bm{\Delta}_\perp\cdot \bm{\rho}] K_1(\Delta_z \rho)\bm\rho/
\rho^{1+2iZ\alpha}\,,\nonumber\\
\bm{A}_B(\bm{\Delta})&=&-2iZ\alpha \Delta_z \int d\bm{\rho}\phi(\bm\rho)
\exp[-i\bm{\Delta}_\perp\cdot \bm{\rho}] K_1(\Delta_z \rho)\bm\rho/ \rho
 \,.
\end{eqnarray}
If $b\gg |\Delta_z|^{-1}\sim\Delta_{min}^{-1}$ then we can simply replace
$\phi(\bm\rho)\to \phi(0)$ in Eq. (\ref{eq:F_Coul}) so that the differential
distribution does not change as compared with the case of a plain wave.
Therefore, we consider the case $b\ll \Delta_{min}^{-1}$, when the finiteness
of the beam size is very important. In this case we can replace
$K_1(\Delta_z\rho)\to (\Delta_z\rho)^{-1}$ in Eq. (\ref{eq:F_Coul}).

\begin{figure}
 \centering\setlength{\unitlength}{0.1cm}
\begin{picture}(85,80)
 \put(40,0){\makebox(0,0)[t]{$\zeta$}}
 \put(-6,30){\rotatebox[origin=c]{90}{$\pi\rho_0^2\Delta_\perp dR/d\Delta_\perp$}}
\put(0,0){\includegraphics[width=80\unitlength,keepaspectratio=true]{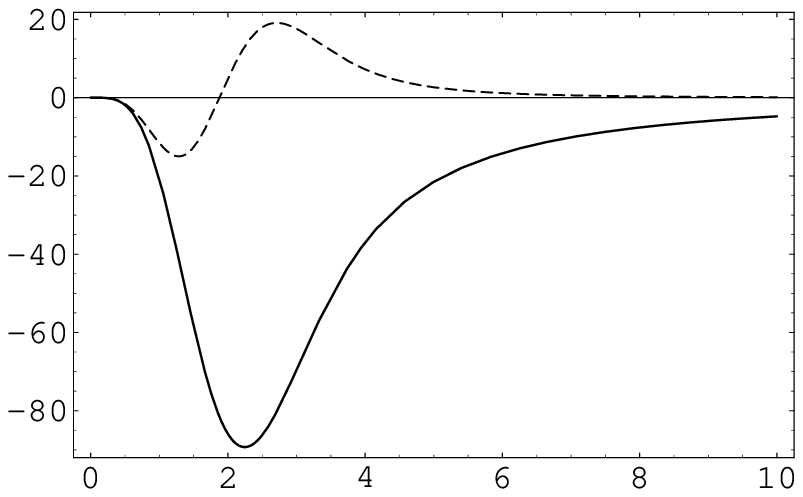}}
\end{picture}
\caption{The quantity $\Delta_\perp\,dR/d\Delta_\perp$ in units
$(\pi\rho_0^2)^{-1}$ as a function of $\zeta=\rho_0\Delta_\perp$ for $Z=80$ and
$\phi(\bm\rho)=\phi_0(\rho)$ (solid curve), $\phi(\bm\rho)=\phi_1(\rho)$
(dashed curve). The functions $\phi_{0,1}$ are defined in Eq.
(\ref{eq:phi01})}\label{fig:FBS}
\end{figure}

Substituting the functions $\bm A(\bm\Delta_\perp)$ and
$\bm{A}_B(\bm\Delta_\perp)$ from Eq. (\ref{eq:F_Coul}) into $dR$ as defined by
Eq. (\ref{eq:dR}) and repeating all the steps of the derivation of $R=\int dR$
in the previous subsection, we obtain
\begin{equation}\label{eq:Rphi}
R= -32 \pi^3(Z\alpha)^2f(Z\alpha)|\phi(0)|^2 \,.
\end{equation}
We see that CC  to the integrated probability depend on the shape of the wave
packet only through the factor $|\phi(0)|^2$, corresponding to the electron
density at zero impact parameter. Thus their dependence on $Z\alpha$ coincides
with that in the case of a plain wave (\ref{eq:R1}). However, the shape of
$\phi(\bm\rho)$ can essentially modify the $\bm\Delta_\perp$-dependence of
$dW_C$. As an illustration, in Fig. \ref{fig:FBS} we show the dependence of
$\Delta_\perp dR/d\Delta_\perp$ on $\zeta$ for $Z=80$ and
$\phi(\bm\rho)=\phi_0(\rho)$ (solid curve) and $\phi(\bm\rho)=\phi_1(\rho)$
(dashed curve), where
\begin{equation}\label{eq:phi01}
\phi_0(\bm\rho)=\frac{\exp[-\rho^2/2\rho_0^2]}{\sqrt{\pi\rho_0^2}} \,, \quad
\phi_1(\bm\rho)=\frac{(\rho/\rho_0)^2\exp[-\rho^2/2\rho_0^2]}{
\sqrt{2\pi\rho_0^2}}\,,\quad \zeta=\rho_0\Delta_\perp\,.
\end{equation}
It is seen that the behavior of $\Delta_\perp dR/d\Delta_\perp$ differs
drastically for the two cases considered. In accordance with Eq.
(\ref{eq:Rphi}), $R=-32\pi^3(Z\alpha)^2f(Z\alpha)/\pi\rho_0^2$ for
$\phi(\bm\rho)=\phi_0(\rho)$ and $R=0$ for $\phi(\bm\rho)=\phi_1(\rho)$. Note
that in the latter case the function $\Delta_\perp dR/d\Delta_\perp$ itself is
different from zero.

\section{Next-to-leading terms in the bremsstrahlung spectrum}
\label{sec:correction_to_spectrum}

As known \cite{Olsen1955} the modification of the high-energy asymptotics of CC
to the spectrum due to the effect of screening is small. Below we show that the
same is true also for the next term in $m/\varepsilon$. In this section we
explicitly calculate the screening correction in the main term of the
high-energy asymptotics and neglect screening when calculating the
next-to-leading term in $m/\varepsilon$. In other words, we calculate the first
corrections in the small parameters $m/\varepsilon$ and $1/mr_{scr}$ to the
bremsstrahlung spectrum
\begin{equation}\label{eq:spectrum1}
\frac{d\sigma^\gamma}{d\omega}=\frac{\alpha\omega
p'\varepsilon'}{2(2\pi)^4}\,\int d\Omega_{\bm{p'}}\,d\Omega_{\bm{k}}
\sum_{\lambda_e\lambda'_e\lambda_\gamma}|M|^{2}\,,
\end{equation}
the amplitude $M$ is given by Eq. (\ref{eq:BS_amplitude_via_WF}) and summation
is performed over the polarizations of all particles. It is convenient to
calculate ${d\sigma^\gamma}/{d\omega}$ using the Green's function
$G(\bm{r}_2,\bm{r}_1|\varepsilon)$ of the Dirac equation in an external field.
This Green's function can be represented as
\begin{eqnarray}\label{eq:FG}
G(\bm r_2 ,\bm r_1|\,\varepsilon)\,&=&\,\sum_{\lambda_en} \frac{\psi_{n}(\bm
r_2)\bar\psi_{n}(\bm r_1 )}
{\varepsilon -\varepsilon_{n}\, + i0} \nonumber\\
&&+ \sum_{\lambda_e}\int \frac{d\bm{p}}{(2\pi)^{3}}\,\left[\,\frac{\psi_P(\bm
r_2 ) \bar\psi_P(\bm r_1 )}{\varepsilon -\varepsilon_{p}\,+ i0}\,+
\,\frac{\psi_{-P}(\bm r_2 )\bar\psi_{-P}(\bm r_1 )} {\varepsilon
+\varepsilon_{p}\,- i0}\,\right] \,\,,
\end{eqnarray}
where $\psi_{n}$ is the discrete-spectrum wave function, $\varepsilon_{n}$ is
the corresponding binding energy, $P=(\varepsilon_{p},\bm p)$. In Eq.
(\ref{eq:FG}) one can use the set of either \emph{in-} or \emph{out-} wave
functions. The regularization of denominators in Eq. (\ref{eq:FG}) corresponds
to the Feynman rule. From Eq. (\ref{eq:FG}),
\begin{eqnarray}\label{eq:FG1}
\sum_{\lambda_e}\int d\Omega_{\bm{p}}\ \psi_P^{(in)}(\bm r_1 )
\bar\psi_P^{(in)}(\bm r_2 )&=&\sum_{\lambda_e}\int d\Omega_{\bm{p}}\
\psi_P^{(out)}(\bm r_1 ) \bar\psi_P^{(out)}(\bm r_2 )= i
\frac{(2\pi)^2}{\varepsilon_{p}p}
 \delta G\,(\bm r_1 ,\bm r_2|\varepsilon_{ p})\,,
\end{eqnarray}
where $\Omega_{\bm{p}}$ is the solid angle of $\bm p$, and $\delta
G=G-\tilde{G}$. The function $\tilde{G}$ is obtained from (\ref{eq:FG}) by the
replacement $i0\leftrightarrow -i0$. Since the spectrum of bremsstrahlung is
independent of the direction of the vector $\bm p$, we can average the
right-hand side of Eq. (\ref{eq:spectrum1}) over the angles of this vector.
Then we obtain, using Eq. (\ref{eq:FG1})

\begin{eqnarray}\label{eq:spectrum}
 \frac{d\sigma^\gamma}{d\omega}=
- \frac{ \alpha\omega}{2\varepsilon p} \int
\frac{d\Omega_{k}}{4\pi}\int\!\!\!\!\int d\bm r_1\, d\bm r_2\,
\mbox{e}^{-i\bm{k}\cdot\bm{r}}\,\sum_{\lambda_\gamma}\,\mathrm{Sp}\,\left\{
 \delta G(\bm r_2,\bm r_1|\varepsilon)\,\hat{e}\,
 \delta G(\bm r_1,\bm r_2|\varepsilon')\,\hat{e}\right\}\,,
\end{eqnarray}
where $\bm r=\bm r_2-\bm r_1$ and $\varepsilon'=\varepsilon-\omega$ is the
energy of the final electron. Here and below we use linear polarization basis
($\bm e^*=\bm e$). Note that the integration over $d\Omega_{\bm k}$ is trivial
since the integrand is independent of  the angles of $\bm k$, so the integral
$\int {d\Omega_{k}}/{4\pi}$ is omitted below. It is convenient to represent
$d\sigma^\gamma/d\omega$ in another form using the Green's function
$D(\bm{r}_2,\bm{r}_1|\varepsilon)$ of the squared Dirac equation,
\begin{equation}\label{eq:FGD}
G(\bm{r}_2,\bm{r}_1 |\varepsilon )= \left[ \gamma^{0}(\varepsilon-
V(\bm{r}_2))-\bm{\gamma}\cdot\bm{p}_2+m \right] D(\bm{r}_2,\bm{r}_1
|\varepsilon )\,,\quad \bm p_2=-i\bm\nabla_2
\end{equation}

Performing transformations as in Refs. \cite{LM1995,LMS2004}, we can rewrite
Eq. (\ref{eq:spectrum}) in the form
\begin{eqnarray}\label{eq:spectrD}
\frac{d\sigma^\gamma}{d\omega}&=-&\frac{\alpha\omega}{4\varepsilon p}
 \int\!\!\!\!\int\!\! d\bm{r}_1 d\bm{r}_2\,\mbox{e}^{-i\bm k\cdot\bm r}\nonumber\\
&& \times\sum_{\lambda_\gamma}\mbox{Sp}\{
 [(2\bm e\cdot\bm p_2-\hat e\hat k)\delta D(\bm r_2 ,\bm r_1 |\varepsilon)]
 [(2\bm e\cdot\bm p_1+\hat e\hat k) \delta D(\bm r_1 ,\bm r_2|\varepsilon')]\}
 \,,
\end{eqnarray}

For the first two terms of the high-energy asymptotic expansion of the
spectrum, the main contribution to the integral in Eqs. (\ref{eq:spectrum}),
(\ref{eq:spectrD}) is given by the region $r=|\bm r_2-\bm r_1|\sim
1/\Delta_{min}= 2\varepsilon\varepsilon'/\omega m^2\gg 1/m$. This estimate is
in accordance with the uncertainty relation. Substituting in Eq.
(\ref{eq:spectrD}) $\delta D$ as $\delta D=D-\tilde D$, we obtain four terms.
Within our accuracy the terms containing $D(\varepsilon)D(\varepsilon')$ and
$\tilde D(\varepsilon)\tilde D(\varepsilon')$ can be omitted and we have
\begin{eqnarray}\label{eq:D_tildeD}
\frac{d\sigma^\gamma}{d\omega}&=&\frac{\alpha\omega}{2\varepsilon p}\textrm{Re}
 \int\!\!\!\!\int\!\! d\bm{r}_1 d\bm{r}_2\,\mbox{e}^{-i\bm k\cdot\bm r}\nonumber\\
&& \times\sum_{\lambda_\gamma}\mbox{Sp}\{
 [(2\bm e\cdot\bm p_2-\hat e\hat k)D(\bm r_2 ,\bm r_1 |\varepsilon)]
 [(2\bm e\cdot\bm p_1+\hat e\hat k) \tilde D(\bm r_1 ,\bm r_2|\varepsilon')]\}
 \,,
\end{eqnarray}
Here and below we assume the subtraction from the integrand of its value at
$Z\alpha=0$. For calculations in the leading approximation in $m/\varepsilon$,
the following form of the function $D(\bm{r}_2,\bm{r}_1 |\varepsilon )$ can be
used \cite{LM1995}
\begin{equation}
D(\bm{r}_2,\bm{r}_1 |\varepsilon )=\left[1+\frac{\bm\alpha\cdot(\bm p_1+\bm
p_2)}{2\varepsilon}\right]D^{(0)}(\bm{r}_2,\bm{r}_1 |\varepsilon)\,,
\label{eq:DviaD0}
\end{equation}
where $D^{(0)}(\bm{r}_2,\bm{r}_1 |\varepsilon )$ is the quasiclassical Green's
function of the Klein-Gordon equation in the external field. The function
$\tilde{D}$ is obtained from Eq. (\ref{eq:DviaD0}) by the replacement
$D^{(0)}\to D^{(0)*}$. The representation (\ref{eq:DviaD0}) can be directly
used for the calculation of the screening correction to the spectrum. It will
be shown below that it can be used for the calculation of the correction in
$m/\varepsilon$ as well.

Substituting Eq. (\ref{eq:DviaD0}) in Eq. (\ref{eq:D_tildeD}) and taking the
trace, we obtain
\begin{eqnarray}\label{eq:D0D0}
&&\frac{d\sigma^\gamma}{d\omega}=
\frac{2\alpha\omega}{\varepsilon^2}\mbox{Re}\int\!\!\!\!\int d\bm r_1 d\bm r_2
\mbox{e}^{-i\bm k\cdot \bm r} \sum_{\lambda_\gamma} \biggl\{4 [\bm e\cdot\bm
p_2 D^{(0)}_2][\bm
e\cdot\bm p_1 D^{(0)}_1]\nonumber\\
&& -\frac{\omega^2}{\varepsilon\varepsilon'}[\bm e\cdot(\bm p_1+\bm p_2)
D^{(0)}_2][\bm e\cdot(\bm p_1+\bm p_2) D^{(0)}_1]\biggr\}\,,\nonumber\\
&&D^{(0)}_2= D^{(0)}(\bm r_2,\bm r_1|\varepsilon)\,,\quad D^{(0)}_1=
D^{(0)*}(\bm r_1,\bm r_2|\varepsilon')\,.
\end{eqnarray}
At the derivation of Eq. (\ref{eq:D0D0}) we integrated by parts the terms
containing second derivatives of $D^{(0)}$. We are interested in CC  which can
be obtained from Eq. (\ref{eq:D0D0}) by the additional subtraction from the
integrand of the Born term ($\propto (Z\alpha)^2$).

\subsection{Next-to-leading term in $m/\varepsilon$ for CC to the spectrum}

We start with Eq. (\ref{eq:D_tildeD}) and introduce the variables
\begin{equation}
\bm r=\bm r_2-\bm r_1,\quad \bm\rho=\frac{\bm r\times[\bm r_1\times\bm
r_2]}{r^2},\quad z=-\frac{(\bm r\cdot\bm r_1)}{r^2}\, .
\end{equation}
Note that the variable $\bm\rho$ in this section has quite different meaning
than the variable $\bm\rho$ in the representation for $\bm{A}(\bm{\Delta})$ in
the previous section, see Eq. (\ref{eq:M_Small_Delta}). The analysis performed
shows that the main contribution to the term under discussion originates from
the region $\rho\sim 1/m$ and $\theta,\psi\sim m/\varepsilon\ll 1$, where
$\theta$ is the angle between the vectors $\bm r_2$ and $-\bm r_1$, and $\psi$
is the angle between the vectors $\bm r$ and $\bm k$. Then screening can be
neglected and we can use the quasiclassical Green's function $D$ in a Coulomb
field obtained in Ref. \cite{LMS2004}
\begin{eqnarray}\label{eq:Dquasiclassical}
&&D(\bm{r}_2,\bm{r}_1 |\varepsilon )=
 \frac{i\kappa\mbox{e}^{i\kappa r}}{8\pi^2r_1r_2}\int d\bm{q}
 \exp\left[i\,\frac{\kappa r q^2 }{2r_1r_2} \right]
 \left(\frac{2\sqrt{r_1r_2}}{|\bm{q}-\bm{\rho}|}\right)^{2i Z\alpha \lambda}
\nonumber\\
 &&\times\left\{
 \left(1+\frac{\lambda r}{2r_1r_2}\bm{\alpha}\cdot \bm{q}\right)
 \left(1+i \frac{\pi(Z\alpha)^2}{2\kappa|\bm q-\bm \rho|}\right)
 -\frac{\pi(Z\alpha)^2}{4\kappa^2}(\gamma^0\lambda-\bm \gamma\cdot \bm r/r)
 \frac{\bm \gamma\cdot (\bm q-\bm \rho)}{|\bm q-\bm\rho|^3}
 \right\}\,,\nonumber\\
&& \lambda=\mathrm{sgn}\,\varepsilon\,,\quad \kappa=\sqrt{\varepsilon^2-m^2}\,,
\quad \bm\alpha=\gamma^0\bm\gamma\, .
\end{eqnarray}

Here $\bm q$ is a two-dimensional vector in the plane perpendicular to $\bm r$.
Note that due to the smallness of the angle $\theta$ we can assume that the
variable $z$ belongs to the interval $(0,1)$ and $r_1=r z$, $r_2=r (1-z)$. The
function $\tilde D$ entering Eq. (\ref{eq:D_tildeD}) is obtained from Eq.
(\ref{eq:Dquasiclassical}) by the replacement $\kappa\to -\kappa$ and
$\lambda\to -\lambda$. The contribution of the last term in braces in Eq.
(\ref{eq:Dquasiclassical}) vanishes after taking the trace in Eq.
(\ref{eq:D_tildeD}). Therefore, this term can be omitted in the problem under
consideration. The remaining terms in Eq. (\ref{eq:Dquasiclassical}) can be
represented in the form (\ref{eq:DviaD0}) with
\begin{equation}
D^{(0)}(\bm{r}_2,\bm{r}_1 |\varepsilon )= \frac{i\kappa\mbox{e}^{i\kappa
r}}{8\pi^2r_1r_2}\int d\bm{q}
 \exp\left[i\,\frac{\kappa r q^2 }{2r_1r_2}\right]
 \left(\frac{2\sqrt{r_1r_2}}{|\bm{q}-\bm{\rho}|}\right)^{2i Z\alpha \lambda}
 \left(1+i \frac{\pi(Z\alpha)^2}{2\kappa|\bm q-\bm \rho|}\right)\,.
\label{eq:D0quasiclassical}
\end{equation}

Then, using the relation
\begin{eqnarray}
(\bm e\cdot\bm p_{1,2})D^{(0)}(\bm{r}_2,\bm{r}_1 |\varepsilon )&=&
\frac{i\kappa^2\mbox{e}^{i\kappa r}}{8\pi^2r_1r_2}\int d\bm{q}
 \exp\left[i\,\frac{\kappa r q^2 }{2r_1r_2}\right]
 \left(\frac{2\sqrt{r_1r_2}}{|\bm{q}-\bm{\rho}|}\right)^{2i Z\alpha \lambda}
\nonumber\\
&&
 \times\left(1+i \frac{\pi(Z\alpha)^2}{2\kappa|\bm q-\bm \rho|}\right)
\left(\mp \frac{\bm e\cdot \bm r}r+\frac{\bm e\cdot\bm q}{r_{1,2}} \right)
 \,,
\label{eq:epD0}
\end{eqnarray}
and passing from the variables $\bm r_{1,2}$ to the variables $\bm r$,
$\bm\rho$, and $z$, we obtain from (\ref{eq:D0D0})
\begin{eqnarray}\label{eq:spectrInit}
&&\frac{d\sigma^\gamma_{C}}{d\omega}=
-\frac{\alpha\omega\varepsilon'}{32\pi^4\varepsilon} \mbox{Re}\int \frac{d\bm
r}{r^5}\int_0^1 \frac{dz}{z^2(1-z)^2} \int\!\!\!\!\int\!\!\!\!\int d\bm q_1d\bm
q_2 d\bm \rho\exp\left[\frac{i\omega r}2\left(
\psi^2+\frac{m^2}{\varepsilon\varepsilon'}\right) +i\frac{\varepsilon
q_1^2-\varepsilon'q_2^2}{2r z(1-z)}\right]\nonumber\\
&& \times
 \left\{\left(\frac{Q_2}{Q_1}\right)^{2iZ\alpha}\!\!\!-1+2(Z\alpha)^2
\ln^2\frac{Q_2}{Q_1}+\frac{i\pi(Z\alpha)^2}2
\left[\left(\frac{Q_2}{Q_1}\right)^{2iZ\alpha}\!\!\!-1\right]
\left(\frac1{\varepsilon Q_1}-\frac1{\varepsilon'Q_2}\right)\right\}\nonumber\\
&& \times\sum_{\lambda_\gamma} \left\{ 4\varepsilon\varepsilon' \left(-\bm
e\cdot\bm r+\frac{\bm e\cdot\bm q_1}{1-z}\right)\left(\bm e\cdot\bm r+\frac{\bm
e\cdot\bm q_2}z\right) +\frac{\omega^2}{z^2(1-z)^2}(\bm e\cdot \bm q_1)(\bm
e\cdot \bm q_2)
 \right\}\,,
\end{eqnarray}
where $Q_{1,2}=|\bm q_{1,2}-\bm\rho|$. The integral over $\bm\rho$ can be taken
with the help of the relations (see Appendix B in \cite{LMS2004})
\begin{eqnarray}\label{eq:f_g}
&& f(Z\alpha)=\frac1{2\pi(Z\alpha)^2q^2} \int d\bm\rho
\left[\left(\frac{Q_2}{Q_1}\right)^{2iZ\alpha}\!\!\!-1+2(Z\alpha)^2
\ln^2\frac{Q_2}{Q_1}\right]=\mbox{Re}[\psi(1+iZ\alpha)+C]\nonumber\\
&&g(Z\alpha)=\frac i{4\pi q} \int \frac{d\bm\rho}{Q_2}
\left[\left(\frac{Q_2}{Q_1}\right)^{2iZ\alpha}\!\!\!-1\right] =
Z\alpha\,\frac{\Gamma(1-iZ\alpha)\Gamma(1/2 +i
Z\alpha)}{\Gamma(1+iZ\alpha)\Gamma(1/2 -i Z\alpha)}\,,
\end{eqnarray}
where $\psi(t)=d \ln \Gamma(t)/dt$, $C=0.577...$ is the Euler constant, $
q=|\bm q_1-\bm q_2|$. Then we perform summation over photon polarization, pass
to the variables $\tilde{\bm q}=\bm q_1+\bm q_2$, $\bm q=\bm q_1-\bm q_2$, and
take all integrals in the following order: $d\Omega_{\bm r}$, $d\tilde{\bm q}$,
$d\bm q$, $dr$, $dz$. The final result for CC  to the bremsstrahlung spectrum
reads

\begin{eqnarray}\label{eq:spectrphot}
&&y\frac{d\sigma_C^\gamma}{dy}=-{4\sigma_0}\biggl[ \left(y^2+\frac43(1-y)
\right)
f(Z\alpha)\nonumber\\
&& -\frac{\pi^3(2-y)m}{8(1-y)\varepsilon} \left(y^2+\frac32 (1-y)\right)
\,\mbox{Re}\, g(Z\alpha) \biggr]\,,\nonumber\\
&& y=\omega/\varepsilon  \,,\quad \sigma_0 =\alpha(Z\alpha)^2/m^2\,.
\end{eqnarray}
\begin{figure}
\centering\setlength{\unitlength}{0.1cm}
\begin{picture}(105,80)
%\graphpaper[1](0,0)(70,50)
 \put(56,0){\makebox(0,0)[t]{$y$}}
 \put(-6,30){\rotatebox[origin=c]{90}{$\sigma_0^{-1}y d\sigma^\gamma_C/dy$}}
\put(0,0){\includegraphics[width=100\unitlength,keepaspectratio=true]{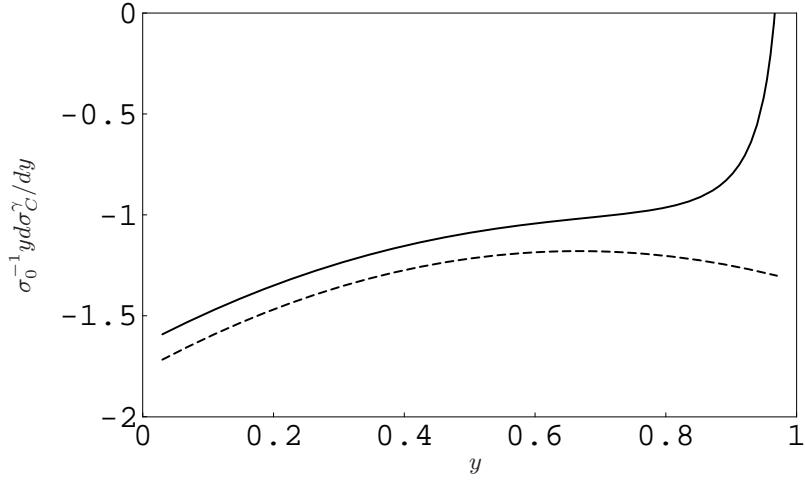}}
\end{picture}
\caption{The dependence of $\sigma_0^{-1} y d\sigma^\gamma_C/dy$ on $y$, see
(\ref{eq:spectrphot}), for $Z=82, \ \varepsilon=50$~MeV. Dashed curve: leading
approximation; solid curve: first correction is taken into account. }
\label{fig1a}\end{figure} In this formula, the term $\propto f(Z\alpha)$
corresponds to the leading approximation \cite{BetheM1954}, the term
$\propto\mbox{Re}\, g(Z\alpha)$ is $O(m/\varepsilon)$-correction. In our recent
paper \cite{LMS2004} this result has been obtained by means of the substitution
rules from the spectrum of pair production by photon in a Coulomb field. The
formula (\ref{eq:spectrphot}) describes bremsstrahlung from electrons. For the
spectrum of photons emitted by positrons, it is necessary to change the sign of
$Z\alpha$ in (\ref{eq:spectrphot}). The $O(m/\varepsilon)$-correction becomes
especially important in the hard part of the spectrum as seen in Fig.
\ref{fig1a}, where $\sigma_0^{-1}y d\sigma^\gamma_C/dy$ with correction (solid
line) and without correction (dashed line) are shown for $Z=82$ and
$\varepsilon= 50$~MeV. Note that in the whole range of $y$ the relative
magnitude of the correction is appreciably larger than $m/\varepsilon$ due to
the presence of large numerical coefficient.

\subsection{Screening corrections} \label{sec:scr_in_spectrum}

In this subsection we calculate the screening correction to the high-energy
asymptotics of $d\sigma_C^\gamma/d\omega$, considering $\lambda_C/r_{scr}$ as a
small parameter.

We start from Eq. (\ref{eq:D0D0}) and use the quasiclassical Green's function
$D^{(0)}(\bm r_2,\bm r_1|\varepsilon)$ for an arbitrary localized potential
$V(\bm r)$. This Green's function has been obtained in \cite{LMS2000} with the
first correction in $m/\varepsilon$ taken into account. The leading term has
the form (see also \cite{LM1995})
\begin{equation}\label{eq:D0V}
D^{(0)}(\bm r_2,\bm r_1 |\,\varepsilon )=\frac{i\kappa\mbox{e}^{i\kappa
r}}{8\pi^2r_1r_2} \int d\bm q \exp\left[i \frac{\kappa
r\,q^2}{2r_1r_2}-i\lambda r\int_0^1dx V\left(\bm r_1+x\bm r -\bm q\right)
\right] \,.
\end{equation}
Similar to Eq. (\ref{eq:spectrInit}) we obtain
\begin{eqnarray}\label{eq:spectr_scr_Init}
&&\frac{d\sigma^\gamma_{C}}{d\omega}=
-\frac{\alpha\omega\varepsilon'}{32\pi^4\varepsilon} \mbox{Re}\int \frac{d\bm
r}{r^5}\int_0^1 \frac{dz}{z^2(1-z)^2} \int\!\!\!\!\int\!\!\!\!\int d\bm q_1d\bm
q_2 d\bm \rho\nonumber\\
&& \times\exp\left[i\Phi+\frac{i\omega r}2\left(
\psi^2+\frac{m^2}{\varepsilon\varepsilon'}\right) +i\frac{\varepsilon
q_1^2-\varepsilon'q_2^2}{2r z(1-z)}\right]\nonumber\\
&& \times\sum_{\lambda_\gamma}\left\{ 4\varepsilon\varepsilon' \left(-\bm
e\cdot\bm r+\frac{\bm e\cdot\bm q_1}{1-z}\right)\left(\bm e\cdot\bm r+\frac{\bm
e\cdot\bm q_2}z\right) +\frac{\omega^2}{z^2(1-z)^2}(\bm e\cdot \bm q_1)(\bm
e\cdot \bm q_2)
 \right\}\,,
\end{eqnarray}
where
\begin{eqnarray}\label{eq:Phi1}
 &&
 \Phi=r \int_0^1 dx[V(\bm r_1+x\bm r -\bm q_2)-V(\bm r_1+x\bm r
-\bm q_1)]\,.
\end{eqnarray}

As we shall see, it is meaningful to retain the screening correction only in
the case $r_{scr}\ll \Delta_{min}^{-1}$, which is considered below. Then the
main contribution to the integral (\ref{eq:spectr_scr_Init}) comes from the
region $1/m\lesssim\rho\lesssim r_{scr}\ll r$ and $q_{1,2}\sim 1/m$. Under
these conditions, the narrow region $\delta x=\rho/r\ll 1$ around the point
$x_0=-\bm r_1\cdot \bm r/r^2=z$ is important in the integration over $x$ in Eq.
(\ref{eq:Phi1}). Therefore, we can perform this integration from $-\infty$ to
$\infty$. After that the phase $\Phi$ becomes
\begin{eqnarray}\label{eq:PhiScr}
\Phi&=&2Z\alpha \ln(Q_2/Q_1)+\Phi^{(scr)}\nonumber\\
&=&2Z\alpha \ln(Q_2/Q_1)+ r \int_{-\infty}^\infty dx[\delta V(\bm r_1+x\bm r
-\bm q_2)-\delta V(\bm r_1+x\bm r -\bm q_1)]\,,
\end{eqnarray}
where $\delta V(\bm r)$ is the difference between an atomic potential and a
Coulomb potential of a nucleus. The notation in Eq. (\ref{eq:spectr_scr_Init})
and in Eq. (\ref{eq:PhiScr}) is the same as in Eq. (\ref{eq:spectrInit}). It is
seen that $\Phi_{scr}\sim \rho\, \delta V(\rho)\sim Z\alpha\, \delta V(\rho)/
V(\rho)\ll 1$ for $\rho\sim m$ and $\Phi_{scr}\sim q_{1,2}/\rho\sim 1/m\rho\ll
1$ for $\rho\sim r_{scr}\gg 1/m$. Thus, expression (\ref{eq:spectr_scr_Init})
can be expanded in $\Phi^{(scr)}$. In our calculation of the screening
correction $d\sigma_C^{\gamma(scr)}/d\omega$, we retain the linear term of
expansion in $\Phi^{(scr)}$. The function $\delta V(\bm R)$ can be expressed
via the atomic electron form factor $F(\bm Q)$ as follows
\begin{equation}\label{eq:FF}
\delta V(\bm R)=\int \frac{d\bm {Q}}{(2\pi)^3}\, e^{i\bm {Q}\cdot\bm
R}\,F(\bm{Q})\frac{4\pi Z\alpha}{{Q}^2}\, .
\end{equation}
Substituting this formula into Eq. (\ref{eq:PhiScr}) and taking the integral
over $x$ from $-\infty$ to $\infty$, we obtain for $\Phi^{(scr)}$
\begin{equation}
\Phi^{(scr)}=\int \frac{d\bm {Q}_\perp}{(2\pi)^2}\, \left(e^{i\bm
{Q}_\perp\cdot(\bm \rho-\bm q_2)}-e^{i\bm {Q}_\perp\cdot(\bm \rho-\bm
q_1)}\right)\,F(\bm{Q}_\perp)\frac{4\pi Z\alpha}{{Q}_\perp^2}\, ,
\end{equation}
where $\bm{Q}_\perp$ is a two-dimensional vector lying in the plane
perpendicular to $\bm r$. Then we use the identity ( see Eqs. (22) and (23) in
\cite{LMS1998a})
\begin{eqnarray}\label{eq:rho2f}
&&\int d\bm \rho \left(\frac{|\bm\rho-\bm q_2|}{|\bm\rho-\bm
q_1|}\right)^{2iZ\alpha}\exp\left[{i \bm{Q}_\perp\cdot (\bm \rho-\bm
q_{1,2})}\right]\nonumber\\
&&= \frac{q^2}{4{Q}_\perp^2}\int d\bm f
\left(\frac{f_2}{f_1}\right)^{2iZ\alpha}\exp\left[{i \bm q\cdot \bm
f_{1,2}/2}\right]\,,
\end{eqnarray}
where $\bm q=\bm q_1-\bm q_2$ and $\bm f_{1,2}=\bm f\mp\bm{Q}_\perp$. Expanding
the exponential function in Eq. (\ref{eq:spectr_scr_Init}) with respect to
$\Phi^{(scr)}$ and using the relation (\ref{eq:rho2f}), we take the integrals
over $\bm q_{1,2}$, $\bm r$, and $z$ and obtain
\begin{eqnarray}\label{eq:scrspect1}
&&y\frac{d\sigma_C^{\gamma(scr)}}{dy}=\frac{4\alpha(Z\alpha)}{\pi}
\mbox{Im}\int \frac{d\bm {Q}_\perp}{{Q}_\perp^4}F(\bm{Q}_\perp)
 \int\frac{d\bm f}{2\pi}\left[\left(\frac{f_2}{f_1}\right)^{2iZ\alpha}-2iZ\alpha \ln\frac{f_2}{f_1}\right]
\left[\frac{S(\xi_1)}{f_1^2}-\frac{S(\xi_2)}{f_2^2} \right]\,,\nonumber\\
 &&S(\mu)=\frac{(\mu-1)}{\mu^{2}}\Biggl\{
 \frac{1}{2\sqrt{\mu}}\left[y^2(3-\mu)+(y-1)(\mu^2+2\mu-3)\right]\,
 \ln\left[\frac{\sqrt{\mu}+1}{\sqrt{\mu}-1}\right]\nonumber\\
 &&-3y^2-(y-1)(\mu-3)\Biggr\}\,,\nonumber\\
 &&y=\omega/\varepsilon\,,\quad\xi_{1,2}=1+16m^2/f_{1,2}^2\,.
\end{eqnarray}

\begin{figure}
\centering\setlength{\unitlength}{0.09cm}
\begin{picture}(105,80)
%\graphpaper[1](0,0)(70,50)
 \put(56,0){\makebox(0,0)[t]{$Z$}}
 \put(-6,30){\rotatebox[origin=c]{90}
 {$f(Z\alpha)^{-1} A_{1,2}$}}
\put(0,0){\includegraphics[width=100\unitlength,keepaspectratio=true]{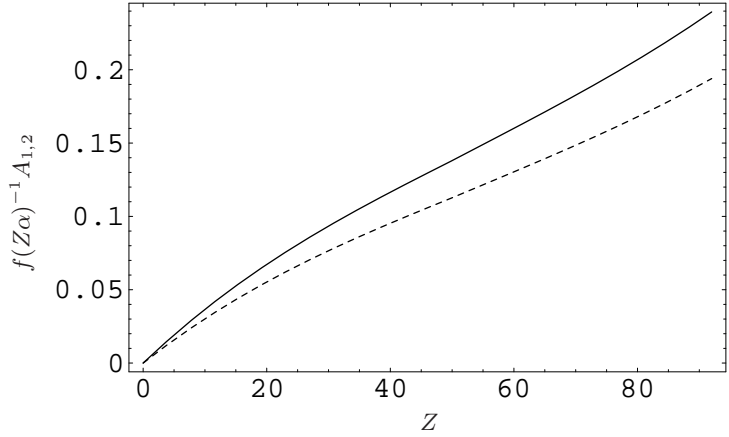}}
\end{picture}
\caption{The dependence of $ A_1/f(Z\alpha)$ (solid curve) and $A_2/f(Z\alpha)$
(dashed curve) on $Z$.} \label{fig:A1A2}
\end{figure}

Using the trick introduced in \cite{LMS1998a}, we can rewrite this formula in
another form. Let us multiply the integrand in (\ref{eq:scrspect1}) by
\begin{eqnarray}
\label{eq:ytrick}
1&\equiv&\int_{-1}^{1}dx\,\delta \left(x-\frac{2\bm f\cdot\bm
{Q}_\perp}{\bm f^2+\bm {Q}_\perp^2} \right)
\nonumber\\
&=&(\bm f^2+\bm {Q}_\perp^2)\int_{-1}^{1}\frac{dx}{|x|} \delta((\bm f-\bm
{Q}_\perp /x)^2 -\bm {Q}_\perp^2(1/x^2-1)) \,,
\end{eqnarray}
change the order of integration over $\bm f$ and $x$, and make the shift $\bm
f\rightarrow \bm f+\bm {Q}_\perp/x$. After that the integration over $f$ can be
easily performed. Then we make the substitution $x=\tanh \tau$ and obtain
\begin{eqnarray}\label{eq:scrspect}
&&y\frac{d\sigma_C^{(scr)}}{dy}=16\sigma_0m^2 \int_0^\infty \frac{d\bm
Q_\perp}{2\pi}\frac{F(\bm Q_\perp)}{Q_\perp^4}\int_0^\infty\frac{d\tau}{\sinh
\tau}\left[\frac{\sin(2Z\alpha\tau)}{2Z\alpha}-\tau\right]\nonumber\\
&&\times\int_0^{2\pi}\frac{d\varphi}{2\pi} \left[e^{\tau}S(\mu_2)-e^{-\tau}
 S(\mu_1)\right] \, ,\nonumber\\
  && \mu_{1,2}=
 1+\frac{8m^2e^{\mp\tau}\sinh^2\tau}{Q_\perp^2(\cosh\tau+\cos\varphi)}\, .
\end{eqnarray}
According to Eq. (\ref{eq:scrspect1}) the correction
$y{d\sigma_C^{\gamma(scr)}}/{dy}$ has the form
\begin{equation}
y\frac{d\sigma_C^{\gamma(scr)}}{dy}=\sigma_0 \left[A_1 (1-y)+A_2 y^2 \right]
\end{equation}
Shown in Fig.~\ref{fig:A1A2} is the $Z$ dependence of the ratio
$A_{1,2}/f(Z\alpha)$ calculated numerically with the use of form factors from
\citep{HO1979}. For the less realistic Yukawa potential, we can perform
analytical calculations of the functions $A_i$. It turns out that their
dependence on the parameter $\beta=\lambda_c/r_{scr}$ has the form
\begin{equation}\label{eq:Ai_Yukawa}
A_i=(Z\alpha)^2\beta^2\left( a_{i} \ln^2\beta +b_{i} \ln \beta +c_{i}
\right)\,,
\end{equation}
where $b_{i}$ and $c_i$ are some functions of $Z\alpha$, while $a_i$ does not
depend on $Z\alpha$. Recollecting that $\beta$ is proportional to $Z^{1/3}$ in
Thomas-Fermi model, we see that $A_i$ depend on $Z$ mainly via the factor
$(Z\alpha)^2\beta^2\propto (Z\alpha)^2Z^{2/3}$. Therefore it is quite natural
that $y{d\sigma_C^{\gamma(scr)}}/{dy}$ calculated with the use of the exact
form factors is well fitted by the following expression
\begin{equation}
y\frac{d\sigma_C^{\gamma(scr)}}{dy}\approx 8.6\cdot 10^{-3}\sigma_0 (Z\alpha)^2
Z^{2/3}[1.2(1-y)+y^2]\,.
\end{equation}
In fact, the accuracy of this fit for all $Z$ is better than a few percent.

It follows from Eq. (\ref{eq:Ai_Yukawa}) that for $r_{scr}\gtrsim
\Delta_{min}^{-1}$ the factor $\beta^2$ in the screening correction is
extremely small, $\beta^2\lesssim (m/\varepsilon)^2$. The terms of such order
were systematically neglected in our consideration. Hence, within our accuracy,
the account of screening correction is meaningful only for $r_{scr}\ll
\Delta_{min}^{-1}$.

\section{Conclusion} \label{sec:Conclusion}

In the present paper we have performed the detailed analysis of CC both to the
differential and the integrated cross sections of bremsstrahlung in an atomic
field. We have calculated the next-to-leading term in the high-energy
asymptotics of the bremsstrahlung spectrum. Similar to the leading term of the
high-energy asymptotics of CC  to the spectrum, this term is independent of
screening in the leading order in the parameter $\lambda_c/r_{scr}$. We have
also calculated the first correction to the spectrum in the parameter
$\lambda_c/r_{scr}$.

We have shown that, in contrast with CC  to the spectrum, CC  to the
differential cross section strongly depend on screening even in the leading
approximation. This dependence is very important in the region giving the main
contribution to the integral over $\Delta_{\perp}$. We have performed the
explicit integration over $\Delta_{\perp}$ of $d\sigma_C^\gamma$ for arbitrary
screening and have verified the independence of the final result on screening.

We also examined the effect of the finite beam size on CC  to bremsstrahlung in
a Coulomb field of a heavy nucleus. Similar to the effect of screening, the
finiteness of the beam size leads to the strong modification of CC to the
differential probability while the probability integrated over $\Delta_\perp$
depends only on the density of the electron beam at zero impact parameter.

This work was supported in part by RFBR Grant No. 03-02-16510 and Russian
Science Support Foundation.

\end{document}